\documentclass[aps, prl, reprint, superscriptaddress]{revtex4-1}
\usepackage{graphicx}
\usepackage{amsmath}
\usepackage{amssymb}
\usepackage{hyperref}
\usepackage{color}
\usepackage{physics}
\usepackage{siunitx}
\usepackage{xcolor}
\usepackage{changes}
\usepackage{comment}
\usepackage{gensymb}
\usepackage{bm}
\usepackage{natbib}
\usepackage{textcomp}
\usepackage{subfigure}
\usepackage{morefloats}

\allowdisplaybreaks
\begin{document}
\title{Controlled symmetry breaking of the Fermi surface in ultracold polar molecules}

\author{Shrestha Biswas}
\author{Sebastian Eppelt}
\author{Weikun Tian}

\affiliation{Max-Planck-Institut f\"{u}r Quantenoptik, 85748 Garching, Germany}
\affiliation{Munich Center for Quantum Science and Technology, 80799 M\"{u}nchen, Germany}

\author{Wei Zhang}
\affiliation{Institute of Theoretical Physics, Chinese Academy of Sciences, Beijing 100190, China} 
\affiliation{School of Physical Sciences, University of Chinese Academy of
Sciences, Beijing 100049, China}

\author{Fulin Deng}
\affiliation{Institute of Theoretical Physics, Chinese Academy of Sciences, Beijing 100190, China} 
\author{Christine Frank}
\affiliation{Max-Planck-Institut f\"{u}r Quantenoptik, 85748 Garching, Germany}
\affiliation{Munich Center for Quantum Science and Technology, 80799 M\"{u}nchen, Germany}

\author{Tao Shi}\email{e-mail: tshi@itp.ac.cn}
\affiliation{Institute of Theoretical Physics, Chinese Academy of Sciences, Beijing 100190, China} 
\affiliation{School of Physical Sciences, University of Chinese Academy of Sciences, Beijing 100049, China}

\author{Immanuel Bloch}
\affiliation{Max-Planck-Institut f\"{u}r Quantenoptik, 85748 Garching, Germany}
\affiliation{Munich Center for Quantum Science and Technology, 80799 M\"{u}nchen, Germany}
\affiliation{Fakult\"{a}t f\"{u}r Physik, Ludwig-Maximilians-Universit\"{a}t, 80799 M\"{u}nchen, Germany}

\author{Xin-Yu~Luo} \email{e-mail: xinyu.luo@mpq.mpg.de}
\affiliation{Max-Planck-Institut f\"{u}r Quantenoptik, 85748 Garching, Germany}
\affiliation{Munich Center for Quantum Science and Technology, 80799 M\"{u}nchen, Germany}

\date{\today}
\begin{abstract}	  
{Long-range anisotropic dipole-dipole interactions between ultracold polar molecules are predicted to drive exotic quantum phases, yet direct many-body signatures of these interactions in degenerate Fermi gases have remained elusive. Here, we report the observation of an interaction-induced controlled deformation of the Fermi surface, providing a clear many-body signature in a deeply degenerate Fermi gas of $^{23}\text{Na}^{40}\text{K}$ molecules. Using double microwave (MW) shielding, we prepare $8 \times 10^3$ molecules at $0.23(1)$ times the Fermi temperature, achieving a three-fold suppression of inelastic losses compared to single MW shielding while preserving strong elastic dipolar scattering. We observe Fermi surface deformations of up to $7\,\%$, more than two times larger than those observed in magnetic atoms, despite operating at two orders of magnitude lower densities. Crucially, we demonstrate continuous tuning of the interaction potential from axial U(1) to biaxial C$_{2}$ symmetry, directly imprinting this geometry onto the Fermi surface. We find excellent agreement between our experimental results and parameter-free Hartree-Fock theory. These results establish MW-shielded polar molecules as a highly tunable platform for exploring strongly correlated dipolar Fermi matter and offer a promising path towards topological superfluidity.
}
\end{abstract}

\maketitle

\section*{Introduction}
Ultracold microwave (MW)-shielded polar molecules are emerging as a powerful platform for engineering long-range interactions and exploring strongly correlated quantum phases~\cite{Schindewolf2025}. This engineered interaction~\cite{Deng2023} yields an effective attraction in the plane of the circular MW polarization, naturally favouring the emergence of a chiral $p_x  + ip_y$ superfluid in a Fermi gas at sufficiently low temperatures ~\cite{Deng2023,Cooper2009, Liu2012, Fedorov2016}, a long-standing goal across various experimental platforms~\cite{Sato2009,Iskin2006, Kallin2012, Ren2019}. Unlike magnetic dipoles or static electric dipoles, which are constrained to cylindrical U(1) symmetry aligned with the external field, MW-shielded interactions offer additional control parameters. 
By adjusting the polarization ellipticity of the shielding field, the effective interaction potential can be continuously tuned from isotropic attraction to highly anisotropic attraction with biaxial C$_2$ symmetry in the plane of the MW. This engineered interaction is predicted to support a chiral $p$-wave superfluid in the U(1) limit~\cite{Cooper2009, Liu2012, Fedorov2016} and drive a topological phase transition to a nematic $p$-wave superfluid as the symmetry is reduced to C$_2$ ~\cite{Read2000, vanLoon2016}. In addition to $p$-wave superfluids, dipolar molecular gases with highly tunable anisotropic attraction are expected to host a rich hierarchy of correlated phases, including stripes, density-waves, and dipolar crystals, prior to instability toward collapse~\cite{Brunn2011, Sieberer2011, Wu2015, Matveeva2012}. 

A fundamental precursor to these interaction-driven phases is the breaking of the Fermi surface's (FS) spherical symmetry and the emergence of momentum-space anisotropy. In a spin-polarized dipolar Fermi gas, mean-field corrections in the $l=2$ angular-momentum channel distort the FS into a quadrupolar shape, elongating it along the direction of strongest attraction.
Analogous FS anisotropy has been observed in strongly correlated solid-state systems, including FS reconstructions in underdoped~\cite{Lalibert2011} and overdoped~\cite{Tam2022} cuprates, and nematic responses in ruthenates~\cite{Borzi2007} and iron-based superconductors~\cite{Bhmer2017}. Despite extensive theoretical predictions~\cite{Miyakawa2008, Shi2010, Sieberer2011, Baranov2012, Baillie2015, Velji2017, Velji2019}, the experimental observation of FSD in ultracold molecular gases has remained elusive, as it requires simultaneously deep degeneracy and strong anisotropic interactions. 

After more than a decade of effort, collisional shielding techniques have enabled ultracold polar molecules to reach quantum degeneracy~\cite{Valtolina2020, Matsuda2020, Anderegg2021, Schindewolf2022, Bigagli2023, Shi2025}. In particular, MW shielding led to the first realization of a degenerate Fermi gas of polar molecules in three dimensions~\cite{Schindewolf2022}. Here, we extend this approach by adding a near orthogonal linearly polarized MW field~\cite{Bigagli2024, Karman2025} to the circular field, suppressing inelastic two-body loss by an additional factor of three and enabling preparation of a deeply degenerate Fermi gas of 8$\times10^3$ $^{23}\text{Na}^{40}\text{K}$ molecules at 0.23(1) times the Fermi temperature. 

In this deeply degenerate regime, we observe a controlled cylindrical symmetry-breaking deformation of the FS of up to $7\, \%$, constituting the first direct observation of interaction-driven reshaping of the FS in a degenerate Fermi gas of polar molecules. Remarkably, despite operating at two orders of magnitude lower densities, the observed deformation is more than two times larger than that reported in a Fermi gas of magnetic atoms~\cite{Aikawa2014}, highlighting the much stronger electric dipolar interactions achievable with molecules. As the MW parameters tune the interaction from a U(1) to a C$_2$  symmetric potential, the FS responds directly to this control by continuously reshaping from an axially symmetric to a biaxially distorted form. Finite-temperature Hartree–Fock theory quantitatively reproduces the observed deformation with no adjustable parameters.

\section*{Fermi Surface Deformation}
\begin{figure}
    \centering
	\includegraphics[width=\linewidth]{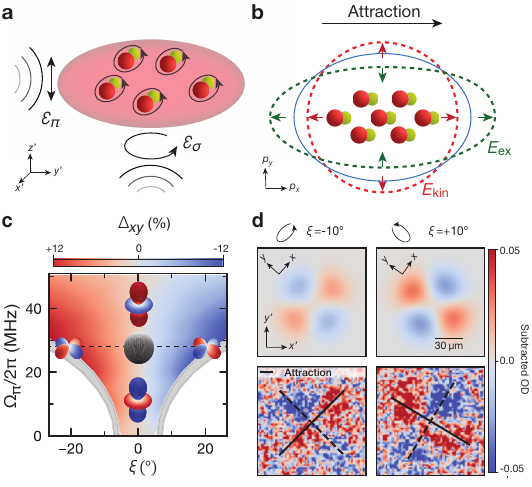}\par
	\caption{\textbf{Fermi Surface Deformation of polar molecules with U(1) and C$_2$  symmetric dipolar interactions.} \textbf{a)} Experimental geometry: Molecules confined in an optical dipole trap are shielded by two MW fields, a circularly polarized field and a nearly orthogonally linearly polarized field.
    \textbf{b)} Mechanism of FSD: Anisotropic Fock exchange energy (green) arising from dipolar interaction elongates the momentum distribution along the direction of attractive interactions, while the isotropic kinetic pressure (red) counteracts it, yielding a squeezed, ellipsoidal FS (blue). \textbf{c)} Hartree–Fock prediction of the dimensionless deformation $\Delta_{xy} = \sigma_x/\sigma_y-1$ as a function of ellipticity $\xi$ and Rabi frequency $\Omega_\pi$ at $T/T_\mathrm{F}=0.2$. Here $\sigma_x$ ($\sigma_y$) denotes the cloud width along $x$ ($y$), the principal axes of the deformed FS. Fixed parameters: $(\Delta_\sigma, \; \Delta_\pi, \;\Omega_\sigma)  = 2\pi\times (20,\, 30,\, 30) \,\mathrm{MHz}$. The dashed line marks the dipolar cancellation point, while shaded gray regions indicate proximity to field-linked resonances~\cite{Chen2022}, where the perturbative theory is no longer valid. \textbf{3D plots} show the dipolar interaction energy landscape at a representative parameter, with attractive (repulsive) interactions shown in red (blue). \textbf{d)} Top: Difference between an elliptical and a circular Fermi–Dirac fit (see \hyperref[sec:image_analysis]{Methods}) to the averaged absorption image (65-70 shots) at $\xi=\pm\ang{10}$. Black ellipses indicate the direction of the MW major axis. Scale bar: $30\,\mu\mathrm{m}$. Bottom: Difference images obtained by subtracting a $\ang{90}$-rotated copy of the cloud from the original image, residuals highlighting the quadrupolar deformation. Black solid (dashed) lines indicate the directions of attractive (repulsive) interactions. $x'-y'$ axes indicate the camera axes, whereas $x-y$ axes indicate the principal axes of the deformed FS.}
	\label{fig:fig_1}
\end{figure}

\begin{figure*}
	\centering
    \includegraphics[width=\linewidth]{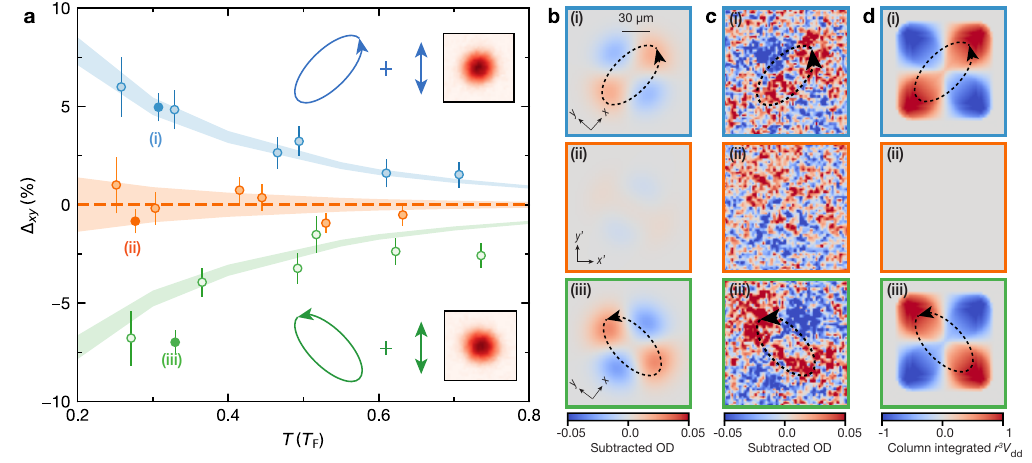}
	\caption{\textbf{Radial Fermi surface deformation vs Fermi degeneracy.} \textbf{a)} Extracted deformation $\Delta_{xy} = \sigma_x/\sigma_y -1$ versus reduced temperature $T/T_{\mathrm{F}}$. Blue, orange, and green circles correspond to $\xi = \ang{-10},\, \ang{0},$ and  $+\ang{10}$, respectively. Error bars are the standard deviation from the fit to the averaged (15–20 shots) images. The shaded areas are finite-temperature Hartree-Fock predictions with no free parameter, taking into account the uncertainty in molecule numbers, $T/T_{\mathrm{F}}$, and $\xi$. \textbf{Insets} Representative absorption images after $14 \, \mathrm{ms}$ of ballistic expansion at $\xi = \pm \ang{10}$ (filled circles). Each panel averages 65-70 shots, each pixel corresponds to $2\times 2$ binning of raw images, with normalized optical depth shown in false colour. \textbf{b)} Residuals from subtracting circular from elliptical Fermi–Dirac fits for the data points marked by filled circles in (a), Scale bar: 30\,$\mu\mathrm{m}$. $x-y$ axes in panels (i) and (iii) show the principal axes of clouds for $\xi = \ang{-10}$ and $+\ang{10}$, respectively.   \textbf{c)} Corresponding self-subtracted images (see text for definition), highlighting the quadrupolar four-lobe pattern.     \textbf{d)} Dimensionless angle dependence of long-range interaction energy, $r^3V_\mathrm{dd}$,  integrated along $z'$ direction for negative, zero and positive $\bar{C}_3$ (top to bottom). In panels (b--d), dashed ellipses show the orientations of MW fields, rows (i), (ii), (iii) correspond to $\xi = -\ang{10}$, $\ang{0}$, $+\ang{10}$ from top to bottom.}
	\label{fig:fig_2}
\end{figure*}

A homogeneous, non-interacting degenerate Fermi gas occupies momentum states within a sphere of radius $k_\mathrm{F} = (6 \pi^2 n_0)^{1/3}$, where $n_0$ is the gas density. At the trap center, our harmonically confined gas is well approximated as quasi-homogeneous under the local density approximation. The anisotropic dipole-dipole interactions (DDI) break the spherical symmetry of the momentum distribution, and the resulting mean-field correction distorts the Fermi sphere while conserving its volume~\cite{Miyakawa2008, Shi2010, Baranov2012, Baillie2015, Velji2017, Velji2019}.

In our system, the dipolar interaction is engineered using two MW fields:  an elliptically polarized MW field $\mathbf{\varepsilon}_\sigma$ and a nearly orthogonal polarized linear field $\mathbf{\varepsilon}_\pi$ (Fig.~\ref{fig:fig_1}a). These fields generate an effective long-range interaction potential between molecules separated by $\mathbf{r}=(r,\theta,\phi)$,
\begin{equation}
    V_\mathrm{dd}(\mathbf{r}) = \frac{1}{r^3} \big[\sqrt{ \frac{16\pi}{5}} C_3 \mathrm{Y}_{2,0}(\theta,\phi) + \sqrt{ \frac{8\pi}{15}}\bar{C}_3  \sum_{m=\pm 2} \mathrm{Y}_{2,m}(\theta,\phi) \big],
    \label{eq:interaction_potential}
\end{equation}
where $C_3$ and $\bar{C}_3$ are tunable interaction coefficients determined by strength, detuning and polarization of the external MW fields~\cite{Wzhang2025}, and $\mathrm{Y}_{2,m}(\theta,\phi)$ are spherical Harmonic functions with $\theta$ denoting the polar angle between $\mathbf{r}$ and $\mathbf{\varepsilon}_\pi$ and $\phi$ denoting the azimuthal angle. The first term represents the standard cylindrically symmetric interaction, while the second term introduces a biaxial anisotropy proportional to $\sin^2\theta \cos(2\phi)$, breaking the U(1) rotational symmetry around the quantization axis. 

In a mean-field description, the dipolar Fermi gas is governed by the interplay of kinetic energy, trap potential, and Hartree–Fock energy arising from long-range anisotropic interactions. Together, these terms determine both the real- and momentum-space density profiles of the gas. The Hartree contribution elongates the real-space density distribution along the direction of strongest attraction (see \hyperref[sec:theory]{Methods}), an effect independent of quantum statistics, as observed in dipolar bosonic gases through magnetostriction~\cite{Stuhler2005} and electrostriction~\cite{Zhang2025, Shi2025} experiments. In dipolar Bose–Einstein condensates (BECs), where particles occupy the same microscopic state, elongation of the real-space wavefunction is accompanied by contraction of the momentum distribution along that direction. In contrast, the Fock exchange term in Fermi gases energetically favours the occupation of momentum states aligned with the axis of strongest attraction, where Pauli repulsion is minimized. This leads to an anisotropic, momentum-dependent self-energy that stretches the FS along the direction of strongest attraction. This alignment of momentum-space and real-space deformations, opposite to the behavior in bosonic systems~\cite{Baillie2012}, serves as a direct many-body signature of dipolar interactions in a degenerate Fermi gas.

The deformation of the FS $\Delta_{xy(\rho z)}$ is defined in the $x-y$ ($\rho-z$) plane, a coordinate system determined by the MW-induced interaction, assuming orthogonal circular and linear MW fields (see \hyperref[sec:coordinate_system]{Methods}). This differs from the laboratory frame used for absorption imaging, denoted as $(x', y',z')$ in Fig.~\ref{fig:fig_1}.  In general, $(x,y,z)$ is rotated with respect to $(x', y',z')$ by an angle that we extract from the fitted orientation of the expanded cloud, which varies under different MW configurations.

\section*{Many-body signature}
\begin{figure*}
    \centering
    \includegraphics[width=\linewidth]{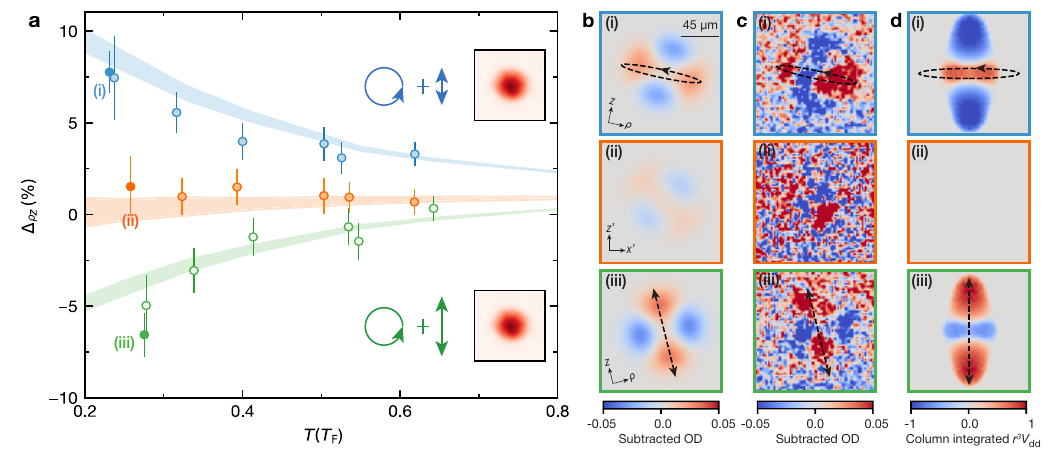}
    \caption{\textbf{Axial Fermi surface deformation vs Fermi degeneracy.} \textbf{a)} Measured deformation $\Delta_{\rho z} = \sigma_\rho/\sigma_z - 1$ versus $T/T_\mathrm{F}$ for $\Omega_\pi = 2\pi\times 15\,\mathrm{MHz}$ (blue, upper), $2\pi\times 30\,\mathrm{MHz}$ (orange, middle), and $2\pi\times 40\,\mathrm{MHz}$ (green, lower). Error bars denote uncertainties from the fits. Shaded bands indicate finite-temperature Hartree–Fock calculations that include trap anisotropy and experimental uncertainties in molecule number, temperature, and $\Omega_\pi$. \textbf{Insets} Representative absorption images of the filled circle data points for $\Omega_\pi \in 2\pi\times (15,\,30,\,40)$\,MHz, each averaged over $60$-$70$ shots, with $2 \times 2$ binning and normalized OD of raw images. \textbf{b)} Residuals obtained by subtracting elliptical and circular Fermi-Dirac fits from the measured distributions for filled circles marked in (a). $\rho-z$ axes in panels (i) and (iii) denote the principal axes in the MW frame when the circular and linear MW fields dominate, respectively. \textbf{c)} Self-subtracted images for the same data. Scale bar: $45 \, \mu\mathrm{m}$. \textbf{d)} Corresponding dimensionless interaction energy $r^3V_\mathrm{dd}$ integrated along $y'$. Ellipses and lines indicate the circular and linear MW field directions, respectively; in panels (b–c), their orientations are inferred from polarization measurements assuming the fields lie in the $x'$–$z'$ plane, while panel (d) shows the ideal case of orthogonal MW fields. %\xyl{The two raw images look the same. Please make sure we use the right raw images.}
    }
    \label{fig:fig_3}
\end{figure*}  

We prepared a deeply degenerate gas of $8\times10^3$ $^{23}\text{Na}^{40}\text{K}$ molecules with a peak density of $2.4(1)\times10^{12}\text{cm}^{-3}$ and at $T = 0.23(1)\, T_\mathrm{F}$, where $T_\mathrm{F}$ is the Fermi temperature of the gas, in a nearly oblate harmonic trap with frequencies $(\omega_x', \omega_y', \omega_z') = 2\pi\times(51,53,140)\,\mathrm{Hz}$. The sample was evaporatively cooled under double MW shielding~\cite{Bigagli2024, Karman2025}, in which nearly orthogonal circularly and linearly polarized MW fields generate a short-range repulsive barrier while preserving a large, tunable long-range dipolar interaction. This scheme strongly suppresses inelastic collisions and reduces the two-body loss coefficient to $2 \times 10^{-13}\,\mathrm{cm}^{3}\mathrm{s}^{-1}$ at $400\, \mathrm{nK}$, a factor of three lower than that with only the circular MW (see~\hyperref[sec:double_MW_shielding]{Methods}). Thereby, we could boost the peak-density of the molecular gas by a factor of 4, which is essential for observing FSD.  

During evaporation, the molecules were shielded with a circularly polarized MW field (Rabi frequency $\Omega_{\sigma^-} = 2\pi\times 30.0(5)~\mathrm{MHz}$, detuning  $\Delta_{\sigma^-} =2\pi\times 20~\mathrm{MHz}$ and polarization ellipticity $\xi = \ang{0.7(2)}$), together with a linearly polarized MW field (Rabi frequency  $\Omega_\pi = 2\pi\times 22.5(6)$ MHz, and detuning $\Delta_{\pi} = 2\pi\times 30$ MHz). We define the polarization ellipticity $\xi$ of the field as $\ang{0}$ for circular and $\ang{45}$ for a linear MW in the $x-y$ plane. After evaporative cooling, we ramped $\xi$ (and thereby $\bar{C}_3$) to the target value over 30\,ms and held the molecules for 20 ms to allow equilibration (see \hyperref[sec:section_mw_grad]{Methods}). During this process, up to $25\,\%$ of the molecules were lost due to inelastic collisions and evaporation, until a new equilibrium was established. To probe the momentum distribution, both MW fields were sequentially switched off over 200~$\mu$s, followed by reverse STIRAP and 14 ms of ballistic expansion. At the end of expansion, a slow magnetic field ramp dissociated the Feshbach molecules into free Na and K atoms, and the density distribution of K was probed by resonant imaging light along the $z'$ axis. 

To visualize the momentum-space anisotropy and identify the orientation of ellipsoidal FS, we analyzed clouds with high deformation using two complementary methods. First, we fitted the image with both an elliptic and a circular Fermi-Dirac function and subtracted the two fits to reveal the residual deformation. Second, the image was rotated by $\ang{90}$ and subtracted from the original one; the resulting patterns were radially integrated to extract the orientation of the major axis (see \hyperref[sec:image_analysis]{Methods}). Both approaches remove the isotropic background and isolate the characteristic four-lobe pattern oriented along the major and minor axes of the interaction potential (Fig.~\ref{fig:fig_2}b,\,c). The deduced angles between the major-axis of the deformed FS and camera $y'$ axis, $\ang{-48(2)}$ ($\ang{56(2)}$) for $\xi=\ang{-10}$ ($\xi=\ang{10}$), are consistent across the two methods. Applying two-dimensional Fermi-Dirac fits (see \hyperref[sec:image_analysis]{Methods}) to the rotated frame aligned with the MW interaction axes yields the two principal half-widths, from which we calculated the dimensionless deformation parameter $\Delta_{xy} =\sigma_x/\sigma_y -1$, where $\sigma_{x}\, (\sigma_y)$ is the half-width of the cloud along $x (y)$ direction. We note that the angle between the major axes at $\xi = +\ang{10}$ and $\xi = \ang{-10}$ is $\sim \ang{104}$ rather than the expected $\ang{90}$. We attribute the $\ang{14}$ discrepancy from orthogonality to the power imbalance and the residual cross-talk between the two MW feeds of the waveguide antenna.

To verify that the observed distortion was a quantum many-body effect, we tracked its dependence on the degree of Fermi degeneracy (quantified by the reduced temperature $T/T_\mathrm{F}$). The temperature dependence of the deformation for $\xi = \ang{0}$ and $\pm\ang{10}$ is shown in Fig.~\ref{fig:fig_2}a, which corresponds to the interactions under U(1) symmetry and C$_2$ symmetry with two near orthogonal axes of attractive interaction, respectively. The deformations at different temperatures for a given $\xi$ were obtained by fitting the cloud in a fixed coordinate system aligned with the principal axes determined at high deformations. For $\xi \approx \ang{0}$ (nearly isotropic interaction in the MW plane), the fitted radii in both coordinates agree within the uncertainty. Upon lowering the temperature, the FS boundary sharpens, and the anisotropic Fock exchange energy increasingly dominates over isotropic kinetic pressure, leading to pronounced cylindrical symmetry breaking of the FS. We incorporated the temperature dependence within Hartree-Fock theory by evaluating the thermal equilibrium phase-space distribution $f(\textbf{r, p}) = 1/(\mathrm{exp}[(\epsilon(\textbf{r,p}) - \mu)/k_\mathrm{B}T] +1) $, where $\epsilon(\textbf{r, p})$ is the quasiparticle energy under dipolar interactions, $\mu$ and $T$ are chemical potential and temperature of the gas, respectively (details in \hyperref[sec:theory]{Methods}). Both experiment and theory show monotonically increasing deformation as $T/T_{\mathrm{F}}$ decreases (Fig.~\ref{fig:fig_2}a), with no indication of the saturation reported for magnetic atoms~\cite{Aikawa2014}.

\section*{Axial Fermi surface deformation}
The FS can also be deformed in the $\rho-z$ plane via the axial DDI ($\mathrm{Y}_{2,0}$ contribution), enabling full three-dimensional control of its shape. The interaction strength can be controlled by tuning $\Omega_\pi$ at a fixed circular MW field. This continuously varies the interaction coefficient $C_3$  from negative to positive, passing through the dipolar cancellation point at $C_3=0$. With the circular MW ellipticity minimized ($\xi = \ang{0.7(2)}$), the interaction potential is almost cylindrically symmetric.

After evaporation in an oblate trap, $\Omega_\pi$ was ramped linearly to target values over 40\,ms, followed by a 50\,ms hold time for equilibration of the molecules. Neither center-of-mass shift nor quadrupolar oscillation of the clouds was observed during the hold time following the 40 ms of ramp (see Fig.~\ref{fig:extended_hold_time} in \hyperref[sec:section_mw_grad]{Methods}). After equilibration, the cloud ballistically expanded for $16~\mathrm{ms}$, and was subsequently imaged with resonant light propagating along $y'$. Due to the cylindrical symmetry of the cloud, line-of-sight integration preserves the ellipsoidal distortion. The column density was fitted with the elliptic Fermi-Dirac profile, and the deformation was quantified by $\Delta_{\rho z}=\sigma_\rho/ \sigma_z -1$. The orientations of the major axes of the deformed FS relative to the camera $z'$ axis were extracted using two complementary methods at the highest deformation, yielding $\ang{75(2)} \;\text{and}\; \ang{13(2)}$ at $\Omega_\pi = 2\pi\times 15$\,MHz and $\Omega_\pi = 2\pi\times 40$\,MHz, respectively. To account for the influence of the oblate trap geometry on the density distribution at finite TOF, we simulated the ballistic expansion of a non-interacting gas in the rotated $x-z$ coordinates and combined this contribution with the deformation predicted by finite-temperature Hartree–Fock theory (see~\hyperref[{sec:tof_analysis}]{Methods}). Measurements spanning $0.25 \le T/T_\mathrm{F} \le 0.65$ show an increasing deformation with decreasing temperature (Fig.~\ref{fig:fig_3}), highlighting that the observed FSD is a genuine quantum many-body effect. The observed orientations of the deformed FS result from finite MW field misalignment and are consistent with the individually measured orientations of the 
$\sigma$ and $\pi$ MW fields of $\ang{12(1)}$ and $\ang{14(1)}$ from the $z'$ axis, respectively (see Fig.\,~\ref{fig:fig_4}a, top right inset). Although polarization measurements do not fix the azimuthal MW angles, agreement between the deduced polar angle of the MW fields and the orientations of the deformed FS indicates that both circular and linear MW fields lie close to the $x'$–$z'$ plane. Because the principal axes of the Fermi surface differ from those of the trap, the evolution of the cloud orientation was measured during expansion and found to be dominated by the momentum-space distribution at 16\,ms TOF (see Fig.~\ref{fig:extended_1}). 

\section*{Deformation control with tunable dipolar length}
\begin{figure}
    \centering
    \includegraphics[width=\linewidth]{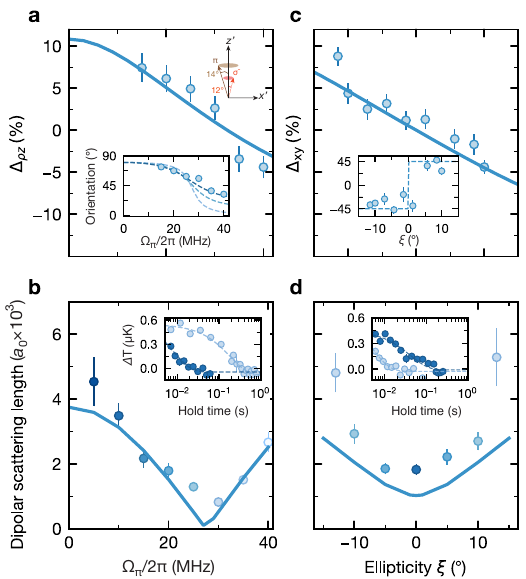}
\caption{\textbf{Control of Fermi surface deformation via tunable dipolar interactions.} \textbf{a)} $\Delta_{\rho z}$ versus $\Omega_\pi$ at $\xi = \ang{0}$. The solid line shows a finite temperature Hartree-Fock calculation at $0.3\,T_\mathrm{F}$, while data span 0.26-0.34 $T_\mathrm{F}$. Error bars denote the fit uncertainties in deformation. \textbf{Bottom inset:} Angle between the cloud major and camera $z'$ axes. Dashed lines indicate the strongest attraction direction for $\ang{30},\, \ang{20}, \, \ang{10}$ (dark to light) misalignment between circular and linear MW field. \textbf{Top right inset} Independent polarization measurements resulting in the normal to circular MW and linear MW fields tilted by $\ang{12}$ and $\ang{14}$ from $z'$, respectively. \textbf{b)} Dipolar length, $a_\mathrm{dd}$ (in units of the Bohr radius $a_0$) from cross-dimensional rethermalization, exhibiting a minimum near the dipolar cancellation point. Error bars include fit uncertainties with density fluctuations of 3–4 runs. Solid lines are $a_\mathrm{dd}$ calculated from coupled channel calculation.  \textbf{Inset:} Raw rethermalization data showing the evolution of temperature difference after directional heating of the cloud at $\Omega_\pi = 2\pi\times 10\,\mathrm{MHz}$ (dark blue) and $\Omega_\pi = 2\pi\times 30\,\mathrm{MHz}$ (light blue), dashed curves are exponential fits. Error bars are the standard error of the mean of 3-4 iterations. \textbf{c)} $\Delta_{xy}$ versus $\xi$ at $\Omega_\pi = 2\pi\times 22.5\,\mathrm{MHz}$. Deformation magnitude grows with $|\bar{C}_3|$ and changes sign at $\xi = \ang{0}$. \textbf{ Inset:} Fitted cloud orientation shifts by $ \sim \ang{90}$ at zero ellipticity, reflecting near-orthogonal rotation of the MW major axis. \textbf{d)} $a_\mathrm{dd}$ versus $\xi$ shows a minimum at $\xi = \ang{0}$, and symmetric growth with $|\xi|$ that follows deformation in (c). \textbf{Inset:} rethermalization traces at $\xi = \ang{0}$ (dark blue) and $\ang{13}$ (light blue).  
}
    \label{fig:fig_4}
\end{figure}

For MW shielded polar molecules, the interaction magnitude and symmetry are continuously tunable through the external MW parameters, enabling a controlled crossover from axial to biaxial geometries. Correspondingly, the FS evolves from an axially symmetric ellipsoid $(a=b\neq c)$ to a fully anisotropic one $(a\neq b\neq c)$, where $a,\;b,\;\text{and}\;c$ denote the principal axes of the FS. This independent control of interaction strength and symmetry allows us to disentangle geometric contributions to FSD across diverse interaction regimes. Here, the dipolar interaction strength is parametrized by the dipolar length $a_\mathrm{dd}$. This sets the separation between two particles at which the dipolar interaction energy equals the zero-temperature kinetic energy and depends on the interaction coefficients $C_3$ and $\bar{C}_3$ via,
\begin{equation}
    a_\mathrm{dd}=\frac{M}{2\hbar^2} \sqrt{C_3^2 + \frac{1}{3} \bar{C}_3^2} ,
\end{equation}
where $M$ is the mass of the dipolar molecules.

We calibrated $a_\mathrm{dd}$ using cross-dimensional rethermalization of low-density thermal molecules at $700\,\mathrm{nK}$, where the kinetic energy is well below the dipolar energy scale (threshold scattering regime)~\cite{Wang2021}. For each MW parameter, we extracted the dipolar scattering length from the cross-dimensional rethermalization time $\tau_{\mathrm{th}}$ as $\sigma_\mathrm{el} =  N_{\mathrm{col}}/(n \tau_{\mathrm{th}} \bar{v})$, where $n$ is the average density, $\bar{v}$ is the average velocity and $N_{\mathrm{col}}$ is the number of collisions required per rethermalization. We assume $\sigma_\mathrm{el}$ to be energy independent, which holds for threshold scattering. Under the Born approximation, $a_\mathrm{dd} =\sqrt{\sigma_\mathrm{el}\times\frac{15}{32\pi}}$ (see \hyperref[sec:section_elastic_sc]{Methods}). At high ellipticities, the assumption of threshold scattering breaks down, leading to discrepancies between the measured and calculated elastic scattering lengths.

Keeping $\Omega_\pi = 2\pi\times 22.5\,\mathrm{MHz}$ fixed and tuning $\xi$ from $\ang{-11.5}$ to $+\ang{10}$, we find that $a_\mathrm{dd}$ is minimized at $\xi = \ang{0}$ and increases three-fold toward either extreme (Fig.~\ref{fig:fig_4}d). 
 For negative $\xi$, the MW major axis is oriented at $\sim \ang{-45}$ from the $y'$ axis, and the deformation decreases approximately linearly as $\xi \to 0$. As $\xi$ becomes positive, the major axis rotates to $\sim \ang{45}$, indicating a near-orthogonal attraction direction, and the deformation again decreases monotonically toward isotropy for $\xi \to 0$. On the other hand, scanning $\Omega_\pi$ from $2\pi\times 5\, \mathrm{MHz}$ to $2\pi\times 40\, \mathrm{MHz}$, we observe that the change of dipolar interaction is asymmetric above and below the dipolar cancellation point, and it grows sharper for  $\Omega_\pi>2\pi\times 30 \; \mathrm{MHz}$. Correspondingly, the cloud deformation along the $z'$ axis increases faster than in the radial direction when the interaction is tuned above the dipolar cancellation point. The principal axes of the deformed cloud are defined by radially integrating the difference between the original image and its $\ang{90}$-rotated counterpart, which reveals a continuous rotation of the major axes from the radial to the axial direction as $\Omega_\pi$ increased. This behaviour arises from higher-order corrections to the effective interaction potential due to the non-orthogonality between the linear and circular MW fields, which introduce a $(\mathrm{Y}_{2,1}+\mathrm{Y}_{2,-1})$ contribution. The inset of Fig.~\ref{fig:fig_4}a shows the best agreement with theory for a relative misalignment of $\ang{20}$–$\ang{30}$, consistent with the individually measured orientations of the MW fields (see~\ref{fig:fig_4}a, inset)\footnote{F.Deng et al. in preparation}. The measured dipolar length $a_\mathrm{dd}$ as a function of $\xi$ is systematically higher than the coupled-channel prediction, which is also attributable to the same non-orthogonality between the two MW fields. 

\section*{Discussion and Conclusion}
In this work, we have observed and characterised the Fermi surface deformation in a deeply degenerate Fermi gas of MW-shielded polar molecules, a fundamental benchmark for realising and exploring strongly-correlated dipolar quantum matter. At a maximum induced lab-frame dipole moment of $0.6\,\mathrm{D}$, we observed deformations as large as $7(1)\,\%$. Remarkably, we achieve this at densities approximately two orders of magnitude lower than in experiments with magnetic atoms~\cite{Aikawa2014}. This difference reflects the much larger molecular dipolar scattering length of $\sim 3000\; a_0$, compared to $\sim 100\; a_0$ for erbium atoms. Despite the reduced density, the ratio of dipolar length to interparticle spacing $a_\mathrm{dd}/d$ increased from 0.038 (magnetic atoms) to 0.21, corresponding to an approximate fivefold enhancement of the dipolar-to-Fermi energy ratio (see~\hyperref[sec:comparisons]{Methods}). This brings the system close to the strongly correlated regime where stripe phases and dipolar crystals~\cite{Brunn2011, Wu2015, Matveeva2012} are expected to emerge due to density-wave instability~\cite{Sieberer2011}.

An ellipsoidally deformed FS concentrates the density of states along the direction of the strongest attractive interaction. This redistribution provides a promising pathway towards the realisation of superfluidity of polar molecules at elevated temperatures by improving pairing along favourable momentum-space directions~\cite{Shi2010}. At temperatures below $0.14\; T_\mathrm{F}$~\cite{Deng2023, Iskin2006}, theory predicts the onset of $p$-wave superfluidity, which should be within experimental reach with increased density or interaction strength. The tunability from axial U(1) to biaxial C$_2$ symmetry provides a route to drive Lifshitz topological transitions between chiral ($p_x + ip_y$) and nematic ($p_x$ or $p_y$) superfluid phases~\cite{Read2000,vanLoon2016}. We note that although similar in-plane attractive interactions can also be engineered for magnetic atoms using rotating magnetic fields~\cite{Giovanazzi2002, Tang2018}, the associated technical heating~\cite{Baillie2020} makes it challenging to maintain the low temperatures required for fermionic superfluid phases.

Beyond equilibrium physics, one can investigate the collective excitation modes of interacting dipolar Fermi systems by applying controlled modulations to the MW fields. Fast modulation of interaction strength can excite quadrupolar motion~\cite{Stringari1996}, while rapid rotation of the interaction axis excites the Scissor mode~\cite{GuryOdelin1999} in the deformed FS. Both collective modes provide sensitive probes of anisotropic hydrodynamics~\cite{Abad2012} and many-body coherence~\cite{Kinast2004, Wright2007}, and may also enable access to multi-particle excitations~\cite{Greiner2005}. These capabilities establish MW-shielded polar molecules as a versatile platform for exploring strongly correlated dipolar Fermi matter, non-equilibrium dynamics, and a promising route towards topological superfluidity.

\section{Acknowledgments}
We thank A. Pelster and A. Balaž for stimulating discussions and sharing their calculations of FSD at zero temperature for microwave shielding with circular MW, X.-Y.Chen, A. Schindewolf, and C. Chin for stimulating discussions. We gratefully acknowledge support from the Max Planck Society, the European Union (PASQuanS grant no. 817482), and the Deutsche Forschungsgemeinschaft under Germany's Excellence Strategy–EXC-2111–390814868  and MCQST seed funding programme. W.T. acknowledges support from the Marie Skłodowska-Curie Postdoctoral Fellowship (grant no. 101207037). T.S. and F.D. acknowledge the NSFC (Grants No.12525413, No.12135018, No.12047503, and No.12504313).

\section*{Author contributions}
All authors contributed substantially to the work presented in this manuscript. S.B. carried out the experiments with assistance from S.E., and W.T., S.B., S.E., and W.T. analyzed the data. S.B, S.E., W.T., and C.F. improved the experimental setup.  W.Z., F.D., and T.S. performed the theoretical calculations. I.B., X.-Y.L., and W.T. supervised the study. All authors contributed to the interpretation of the data and to the final manuscript. 

\section{Data availability}
The experimental data that support the findings of this study are available from the corresponding authors upon request.

\section{Code availability}
All relevant codes are available from the corresponding authors upon request.

\section{Competing interests}
The authors declare no competing interests.

\renewcommand{\bibnumfmt}[1]{#1.}
\bibliography{bibliography}

\section*{Methods}
\subsection{Preparation and detection}
\subsubsection{Ground state molecules}
We prepared a deeply degenerate Fermi gas of $^{23}\text{Na}^{40}\text{K}$ molecules ($8\times 10^3$ at $T = 0.23(1)\,T_\mathrm{F}$) starting from a degenerate mixture of bosonic $^{23}\text{Na}$ and fermionic $^{40}\text{K}$ atoms. First, we prepared loosely bound Feshbach molecules by ramping the magnetic field through the inter-species Feshbach resonance at $78.3\; \mathrm{G}$. These molecules were transferred to deeply bound molecules in their ro-vibrational ground state with a body frame dipole moment of $2.7\; \mathrm{D}$ through STIRAP transfer. To obtain a tunable laboratory-frame dipole moment, we dressed the molecules with MW fields blue-detuned from the transition between the ground state and the first rotationally-excited state of opposite parity. The resulting dipole moment depends on the Rabi frequency and detuning of the dressing MW fields. 

\subsubsection{Trap anisotropy}
\label{sec:tof_analysis}
\begin{figure}
    \centering
    \includegraphics[width=\linewidth]{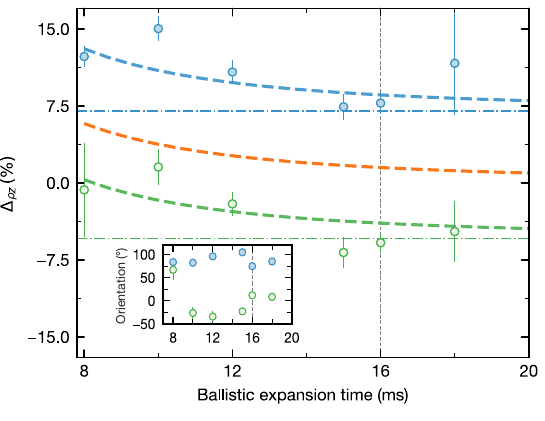}
    \caption{\textbf{Time-of-flight dependence of the axial deformation.} $\Delta_{\rho z}$ measured as a function of ballistic expansion time for two linear-field strengths: $\Omega_\pi = 2\pi\times 15\; \mathrm{MHz}$ (blue circles) and $\Omega_\pi = 2\pi\times 40\; \mathrm{MHz}$ (green circles). The orange-dashed line shows the expected evolution for a non-interacting, isotropic Fermi gas released from the same oblate trap. Blue and green dashed lines are finite-temperature Hartree–Fock predictions that include both the intrinsic momentum-space deformation and the residual real-space anisotropy imposed by the trap and dipole interactions. The dot-dashed line shows the corresponding value of only momentum-space deformation achievable at a long time of flight. Error bars are the standard fitting error of 8-10 averaged images per point. The low density of the cloud resulted in large fitting uncertainties at 18 ms data. \textbf{Inset:} Time-of-flight evolution of the major axis orientation of the deformed cloud. At short times, the orientation is set by the trap axes, while at long times it aligns with the direction of the strongest dipolar attraction. The figures illustrate that both the aspect ratio and the reorientation of the cloud are momentum-space dominated at 16 ms, as shown in Figs.~\ref{fig:fig_3} and ~\ref{fig:fig_4}.}
    \label{fig:extended_1}
\end{figure}
Ground-state molecules were confined in a crossed optical dipole trap made from three 1064-nm beams. Two horizontal beams intersect in the $x'$–$y'$ plane at an angle of $\sim \ang{70}$; their non-orthogonality limits the radial symmetry to $\omega_x'/\omega_y' \simeq 1.2$. To minimise spurious deformation at finite TOF images, we added a vertically propagating beam with cylindrical symmetry about $z'$. With this additional confinement, the horizontal anisotropy was reduced to \(\omega_x'/\omega_y'<1.04\). For measurements in the \(\rho'\!-\!z'\) plane, oblate geometry of the trap with $\frac{\omega_z'}{{\omega_\rho'}}=
\frac{\omega_z'}{\sqrt{\omega_x'\omega_y'}}\simeq 2.6$ resulted in residual deformations in the projected density. We take this effect into account by multiplying the Hartree–Fock momentum-space deformation by the geometric factor, $ \big (\sigma_x /\sigma_z \big)_\textit{real}= \sqrt{(1/\omega_{x}^2 + t_{\mathrm{TOF}}^2)/(1/\omega_{z}^2 + t_{\mathrm{TOF}}^2)}$. The trapping frequencies in the $x-z$ plane are calculated from the $x'-z'$ coordinates following $\omega_{x,z} = \sqrt{\omega_{x',z'}^2\text{cos}^2\varphi+\omega_{z',x'}^2\text{sin}^2\varphi}$, with $\varphi$ being the angle between the two coordinates. Here, \(\sigma_i\) denotes the $1/e$ half-width of the expanded cloud along axis \(i\), and \(t_{\mathrm{TOF}}\) is the free-flight time before absorption imaging (Fig.~\ref{fig:extended_1}).

\subsubsection{Detection of molecules}
Once the sample equilibrated under a certain interaction, the MW fields were adiabatically ramped down in $200~\mu\mathrm{s}$, returning the molecules to the ground state with a vanishing laboratory-frame dipole moment. The optical traps were switched off immediately to prevent distortion of the cloud caused by quenching the MW Rabi frequencies. A reverse-STIRAP pulse converted the ground-state dimers back to the loosely bound Feshbach state, after which the cloud expanded ballistically. During the final \(7.5~\mathrm{ms}\) of time of flight, we slowly ramped the magnetic field across the Feshbach resonance, dissociating the dimers into free Na and K atoms. Absorption images were recorded by probing the K cloud with resonant imaging light and imaging systems providing a calibrated magnification of \(M_{x'y'}=2.33\) and $M_{x'z'}=1.98$.

To verify that the molecule-to-atom conversion does not distort the momentum distribution, we analyzed clouds prepared at the near dipolar cancellation point, where deformation of the FS is expected to be negligible. The extracted deformations, \(0.5(5) \%\) and \(2.2(8) \%\), are consistent with the expected isotropic momentum distribution from the oblate trap at finite time-of-flight.

\subsection{Technical details on the MW fields}
\label{sec:dual_mw_field}
\subsubsection{Polarization characterisation of the circular MW}
\begin{figure}
    \centering
    \includegraphics[width=\linewidth]{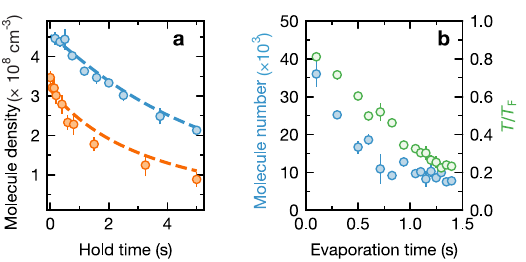}
    \caption{\textbf{Performance of double MW shielding.} \textbf{a)}Molecular density as a function of hold time for samples shielded only by $\sigma$ polarized MW [$\Omega_{\sigma}, \Delta_{\sigma} = 2\pi\times (20,\;20)$ MHz, in orange] and for $\sigma $ + $\pi$ MW  shielding [$\Omega_{\sigma}, \Omega_\pi, \Delta_{\sigma}, \Delta_{\pi} = 2\pi\times (30,\, 22.5, \, 20,\, 30)$ MHz, in blue]. The double-field configuration suppresses inelastic losses, yielding a three-fold reduction of the loss coefficient to $2.3(1)\times 10^{-13}\text{cm}^3\text{s}^{-1}$, compared to $7.4(1)\times 10^{-13}\text{cm}^3\text{s}^{-1}$ for $\sigma$ only shielding. \textbf{b)}Evolution of molecule number and reduced temperature during forced evaporation. Starting from $3.6\times10^4$ molecules at $0.81(1)\; T_\mathrm{F}$, the sample was evaporated to $8\times10^3$ molecules at $T/T_\mathrm{F} = 0.23(1)$. 
    }
    \label{fig:mw_shielding}
\end{figure}

The FS deformation elongates the momentum distribution along the direction of the strongest attraction. When the relative strength of the linear (\(\pi\)) and circular (\(\sigma\)) MW fields was varied, the principal axis of the effective dipole rotates correspondingly, and our data in Secs.~III–IV confirm that the observed deformation follows this axis. For completeness, we verified experimentally that the cloud indeed elongates along the major axis of the MW field, also for $\pm \xi$.

The circular polarization was produced by combining two orthogonal, linearly polarized MW fields with a controlled phase offset. Varying this phase rotates the major axis of the resulting elliptic field by \(\pm\ang{90}\); the sense of the rotation depends on whether the dominant circular component was \(\sigma^{-}\) or \(\sigma^{+}\). From the geometry of our setup, the measured FS elongation shown in Fig.~\ref{fig:fig_1} aligns with the calculated major axis when the circular component was predominantly \(\sigma^{-}\). To quantify the polarization, we measured the Rabi frequencies for driving the transitions from the $|J=0, m_J =0 \rangle$  state to different $m_J$ levels in the first rotational excited state, $|J=1, m_J \in \{-1, 0, +1\} \rangle$. Because the Zeeman splitting between the \(m_{J}\) sub-levels was only $40-100 \,\mathrm{kHz}$ at the magnetic field used ($135 \, \mathrm{G}$), we removed the band-pass filters from the MW path, set the detuning to zero, and attenuated the field such that the single-photon Rabi frequencies are \(\sim 10\;\mathrm{kHz}\)~\cite{Biswas2025}. We also used independently measured relative phase shifts of the removed filters at the operating frequency and corrected the driving phase by $\ang{5}$. The extracted Rabi frequencies are  
$ \Omega_\pi,\,\Omega_{\sigma^{+}},\,
\Omega_{\sigma^{-}} = 2\pi \times [1.56(5), 0.82(2), 9.0(2)]\;\mathrm{kHz}$ confirming a predominantly \(\sigma^{-}\) circular polarization with small \(\sigma^{+}\) and \(\pi\) admixtures that are consistent with the ellipticities used in the main experiment.
\subsubsection{Dual–colour MW shielding for fermionic molecules}\label{sec:double_MW_shielding}

\begin{figure}
    \centering
    \includegraphics[width=\linewidth]{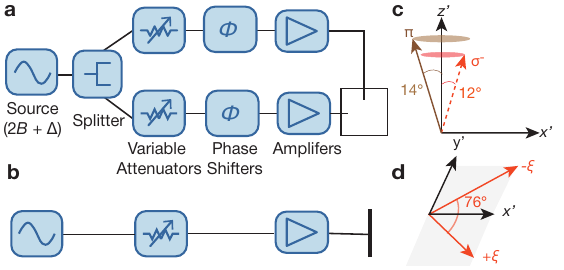}
    \caption{\textbf{Control electronics and Geometry of the MW setup.} \textbf{a)} Electronics used to generate a circularly polarized MW field. By controlling the relative power and phase between the two feeds, we can generate arbitrary polarization ellipticity in the MW. Details of the MW control electronics for the $\sigma$-field can be found in ~\cite{Biswas2025}. \textbf{b)} Electronics used to generate a linearly polarized MW field. We tuned the MW power with the variable attenuator to change the Rabi frequency. \textbf{c)} Orientation of MW fields at the location of molecules. The coordinate axes correspond to the camera frame, with the $ z$-axis pointing in the vertical direction. The normal to the plane of the circularly polarized MW field made an angle of $\ang{12(1)}$ (shown in orange), while the linearly polarized field made an angle of $\ang{14(1)}$ (shown in brown). \textbf{d)} The MW major axes of circular MW for positive and negative ellipticity are shown in dark orange. They are separated by $\ang{104}$ (same as $\ang{76}$ because of C$_2$ symmetry), deviating from the ideal orthogonal configuration.}
    \label{fig:mw_geometry}
\end{figure}

A single, circular MW can suppress the universal short-range loss of ground-state molecules by creating a repulsive shielding barrier \cite{Karman2018, Lassablire2018, Schindewolf2022}. But fermions still suffer from enhanced loss in the vicinity of field-linked (FL) resonances \cite{Chen2022}. Unlike the bosonic case, where evaporation was nearly impossible without additional linear MW fields \cite{Bigagli2023, Lin2023}, for fermionic $^{23}\text{Na}^{40}\text{K}$ molecules, adding a linearly polarized MW was mainly useful to push the resonances to higher ellipticity values at high $\Omega_{\sigma}$. We quantified two-body inelastic loss under circular and double MW shielding by holding the molecular sample for variable durations and fitting the resulting density decay to a two-body loss model~\cite{Biswas2025}. For circular-only MW shielding, we obtained best two-body loss coefficient of $7.4(1)\times 10^{-13}\mathrm{cm}^3\mathrm{s}^{-1}$ at $400\;\mathrm{nK}$, achieved with MW parameters $\Omega_{\sigma}, \Delta_{\sigma} = 2\pi\times (20,\;20)~\mathrm{ MHz}$. In comparison, double MW shielding yields a significantly reduced loss rate of $2.3(1)\times 10^{-13}\mathrm{cm}^3\mathrm{s}^{-1}$ at the same temperature using $\Omega_{\sigma}, \Omega_\pi, \Delta_{\sigma}, \Delta_{\pi} = 2\pi\times (30,\, 22.5, \, 20,\, 30)~\mathrm{MHz}$. Because the linear field can cancel the dipolar effect of the circular MW at long range, both inelastic and elastic scattering were greatly reduced at the dipolar cancellation point. For a fermionic gas, where \(s\)-wave scattering is absent, evaporation would then stall. We therefore operated at slightly below the cancellation condition ($\Omega_\pi =2\pi\times 28\;\mathrm{MHz}$), with an optimized elastic to inelastic scattering rate.

The geometry of MW fields in our setup, as shown in Fig.~\ref{fig:mw_geometry}, were such that the normal to the circular polarization plane made an angle of $\ang{12}$ and the linear field made an angle of $\ang{14(1)}$ from the $z'$ axis, as inferred from the relative strengths of the $\sigma^+$, $\sigma^-$, and $\pi$ polarization components of each field. While these measurements do not reveal the azimuthal orientation, the FSD in the $x'$–$z'$ plane indicates the two fields lie close to the $x'-z'$ plane. This non-orthogonality reduces the efficiency of the double shielding scheme. A fully orthogonal geometry will be implemented in future upgrades and will allow even more precise control of the dipolar interaction tensor.

\subsubsection{MW field gradient}\label{sec:section_mw_grad}
\begin{figure*}
    \centering
    \includegraphics[width=\linewidth]{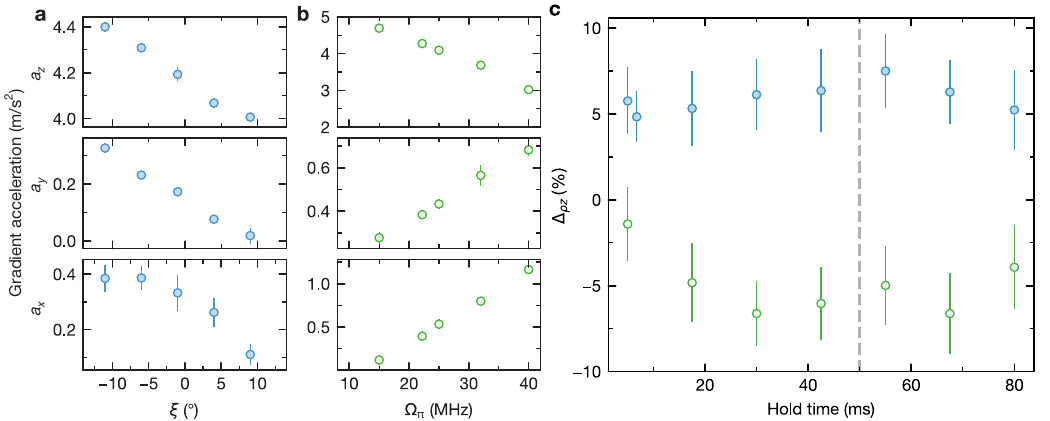}
    \caption{\textbf{Characterization of MW field gradient and deformation versus hold time.} \textbf{a)} and \textbf{b)} Acceleration of untrapped molecules by MW gradient in $x',\; y'\; \text{and}\;z'$ directions as a function of $\xi$ and $\Omega_\pi$. We observe that the change of gradient is more sensitive to $\Omega_\pi$ than to $\xi$. \textbf{c)} Deformation as a function of hold time for $\Omega_\pi =2\pi\times  15\;\mathrm{MHz}$ (blue circles) and $2\pi\times 40\;\mathrm{MHz}$ (green circles). Error bars denote the standard fitting error, with images averaged over 8–10 shots. The dashed line at $50\;\mathrm{ms}$ shows the condition for measurements in the main text.}
    \label{fig:extended_hold_time}
\end{figure*}
The shielding field wavelength ($\lambda_{\mathrm{MW}}\simeq 5.2\;\mathrm{cm}$) is orders of magnitude larger than the typical cloud size ($\sim 20\;\mu\mathrm{m}$), so spatial variations of the field over the sample appear as a linear intensity gradient rather than higher-order curvature. Such gradients arise from the intrinsic emission pattern of the dipole antenna and from reflections off the surrounding metallic hardware.

We quantified the gradient by monitoring the center-of-mass acceleration of a freely expanding cloud when the MWs were left on during expansion time; the acceleration is proportional to the local intensity variation. Across the entire range of $\xi$ and $\Omega_\pi$ used in this paper, the dominant gradient component points along the $z'$-axis, with a magnitude that decreases for positive $\xi$ and for larger $\Omega_\pi$. This gradient depends more sensitively on $\Omega_\pi$ than on $\xi$, as shown in Fig.~\ref{fig:extended_hold_time}a, b. Accordingly, a slower ramp and longer hold were used for $\Omega_\pi$ than for $\xi$ to minimize disturbance on the cloud during the ramp. No center-of-mass oscillations were observed during the hold time (see Fig.~\ref{fig:extended_hold_time}). Instead, we noted a deformation in the momentum distribution, which gradually reduced over extended hold times. This reduction can be attributed to heating caused by collisional losses at long hold times. The shift of the $x',\,y'$ trapping frequencies never exceeds $2\,\%$, and the $z'$ trapping frequency shifts at most $10\,\%$. These variations translate into a 3–5\,\% systematic uncertainty in the extracted reduced temperature $T/T_\mathrm{F}$ and to an uncertainty below $0.5\,\%$  on measured deformation (see Figs.~\ref{fig:fig_2} and \ref{fig:fig_3}).

\subsubsection{Elastic scattering rate}\label{sec:section_elastic_sc}
We determined the elastic scattering cross-section $\sigma_\mathrm{el}$ for a set of MW parameters using cross-dimensional rethermalization of a thermal molecular cloud at $T\simeq 700\;\mathrm{nK}$. Parametric excitation obtained by modulating the trap depth at twice the $z'$ frequency raises the temperature along $z'$ while leaving the radial temperatures almost unchanged. The subsequent relaxation of the temperature imbalance was fitted with $\Delta T = \Delta T_0 e^{-t/\tau_{\mathrm{th}}}$, where $\Delta T = T_{z'} - T_{x'}$ and $\tau_{\mathrm{th}}$ was the rethermalization time. Using a thermal, low-density sample for this measurement ensures $\tau_{\mathrm{th}}$ does not get saturated by hydrodynamic flow~\cite{Ma2003}.

In the collision picture, $\tau_{\mathrm{th}}$ is related to $\sigma_\mathrm{el}$ through $1/\tau_{\mathrm{th}} = n\sigma_\mathrm{el}\bar{v}/N_{\mathrm{col}}$, where $N_{\mathrm{col}}$ is the average number of elastic collisions required per rethermalization and $n$ is the average density and $\bar{v}$ is the average velocity of particles. For collision energy much lower than the dipole-dipole energy (threshold scattering), $N_{\mathrm{col}}$ is energy-independent~\cite{Wang2021}, and if the equilibration is dominated by short-time dynamics, it is given by,
\begin{equation}
    N_{\mathrm{col}}(\varphi) = \frac{112}{45 + 4\cos(2\varphi) - 17\cos(4\varphi)}
\end{equation}
where $\phi$ is the angle between dipoles and the trap axes. With $\varphi = \ang{12}$ (dominant $\sigma^-$ field, below the dipolar cancellation point), we obtain $N_{\mathrm{col}} = 3.0$; above cancellation $\varphi \simeq \ang{14}$ gives $N_{\mathrm{col}} \approx 2.9$ (dominant $\pi$). We calculate the dipolar interaction strength parameters $C_3$ and $\bar{C}_3$ from coupled-channels calculations for double MW shielding at finite ellipticities. This allows us to obtain the dipolar scattering cross section,
\begin{equation}
    \sigma_\mathrm{el} = \frac{32\pi}{15}\left(\frac{M}{2\hbar^2}\right)^2\left[C_3^2 + \frac{1}{3}\bar{C}_3^2\right] = \frac{32\pi}{15}a_\mathrm{dd}^2.
\end{equation}
This definition generalizes the standard dipolar length to biaxial potentials by incoherently summing the scattering contributions from the axial ($C_3$) and non-axial ($\bar{C}_3$) channels under the Born approximation. The calculation agrees well with the experimentally extracted dipolar length $a_\mathrm{dd}$ when the energy-independent $N_{\mathrm{col}}$ is applied.

\subsubsection{Approximations on coordinate systems}\label{sec:coordinate_system}

The MW-engineered dipolar interaction is formulated in the MW frame, where the z-axis is defined by the $\pi$ field and the $x-y$ plane by the $\sigma$ field, assuming orthogonality between the two fields,  as shown in Eq.~\ref{eq:interaction_potential}. 

Experimentally, measurements are performed in the laboratory frame $(x', y', z')$, with $y'$ as the horizontal and $z'$ as the vertical imaging direction. In the ideal case, $z$ and $z'$ coincide. However, due to imperfect alignment and cross-coupling between antennas, the principal axes of the MW fields deviate from the laboratory axes, which can vary with MW parameters. To connect experiment to theory, we adopt the following approximations: In Fig.~\ref{fig:fig_2}, we measure the projection of FSD onto the $x'-y'$ plane, neglecting the $\ang{12}$ tilt between the $x-y$ plane and the $x'-y'$ plane. In Fig.~\ref{fig:fig_3}, we measure the projection of FSD onto the $x'-z'$ plane, assuming the $z$-axis is in the $x'-z'$ plane. For each set of MW parameters, we determine the effective $x$-axis ($z$-axis) direction in the image plane by fitting the principal axes of the deformed cloud. While calculating the dependence of deformation on temperature in Fig.\,~\ref{fig:fig_2} and \,~\ref{fig:fig_3}, we fix the principal axes for higher-temperature images to those of a most deformed cloud, to improve the robustness of fits at lower deformation by eliminating one free fit parameter. This analysis procedure accounts for the MW-parameter-dependent orientation of the MW polarization axes.

\subsection{Analysing images}\label{sec:image_analysis}
\subsubsection{Benchmarking imaging system}
Two identical Manta G-145 NIR cameras were used to image the cloud in the $x'$-$y'$ plane and the $x'$-$z'$ plane. The typical optical depth of an individual image was $\simeq 0.2$. For each data point in Figs.~\ref{fig:fig_2}, \ref{fig:fig_3}, and \ref{fig:fig_4}, we averaged 15-20 images, except for the highlighted solid point showing the four-lobe pattern. In that case, we averaged 60–70 images to suppress background artefacts. We averaged individual shots after centering them to a common center, determined with a Gaussian fit, then fit the mean image with the two-dimensional Fermi–Dirac profile described below. The quoted statistical error is the square root of the corresponding diagonal element of the fit covariance matrix.

\begin{figure}
    \centering
    \includegraphics[width=\linewidth]{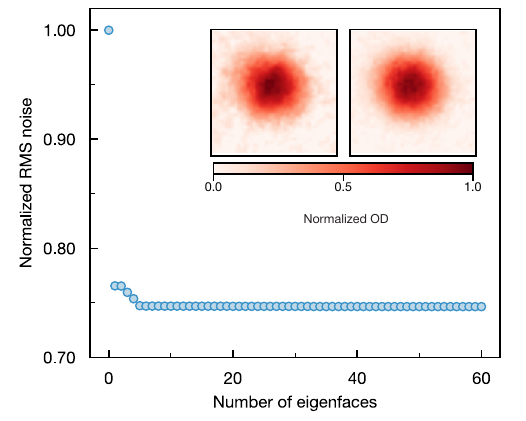}
    \caption{\textbf{Performance of image denoising with the eigenface method.} The correlated noise in the background as a function of the number of eigenfaces used to reconstruct the reference. The eigenfaces were sorted according to their overlap with the image. \textbf{Inset:} The left is the original averaged image. The right is the image after subtracting the reference reconstructed with 60 eigenfaces. The color bar shows a false-color representation of the normalized optical density. 
    }
    \label{fig:extended_eigenface}
\end{figure}

Since the cloud deviates only weakly from a circular shape, we removed spatially correlated background structure before fitting the two-dimensional Fermi–Dirac profile utilizing the background eigenfaces~\cite{Li2007}. About 80–100 reference frames without atoms were reshaped into column vectors and stacked to form a data matrix. We built the covariance matrix from it and computed its eigenvectors, each of which represents a dominant background mode. After sorting the components by descending eigenvalues, we projected each absorption image onto the leading background modes to obtain their weights and reconstruct the background via a weighted sum of eigenvectors. By subtracting the reconstructed background from the absorption image, we suppressed spatially correlated noise. To quantify denoising performance, we evaluated the root-mean-square (RMS) noise in the optical density (OD) values across the field of view for each image, excluding the atomic cloud using a circular mask with radius equal to the Gaussian half-width. We plotted the reduction in background noise relative to no background subtraction as a function of the number of eigenfaces included in the background reconstruction. We observe that the noise decreases sharply for the first six eigenfaces, the first eigenface alone reducing the noise by 24\% (see Fig.~\ref{fig:extended_eigenface}, main panel). The inset shows a representative elliptic cloud before and after denoising within a $90\times 90$ pixel region centered on the cloud.

\subsubsection{Image analysis}
For each data set, we located the principal axes using two complementary methods at high deformation of the cloud. First, we fitted the image with a two-dimensional elliptic Fermi–Dirac distribution that allows for different widths along the principal axes,
\begin{equation}
    \mathrm{OD}(x', y') = A\,\mathrm{Li}_2\left[-\zeta \exp\left\{-\frac{(x'- x_0')^2}{2\sigma_{x'}^2} - \frac{(y'- y_0')^2}{2\sigma_{y'}^2}\right\}\right],
    \label{equation:fermi_dirac}
\end{equation}
where $(x', y')$ are coordinates rotated by an angle $\varphi$ which can also be a fit parameter; $(x_0',y_0')$ denotes the center of the cloud, $\sigma_{x'}$ and $\sigma_{y'}$ are the Gaussian half-widths after expansion, $\zeta$ is the fugacity  and $\mathrm{Li}_2$ is the polylogarithmic function of order two. We then fitted the same data with the constraints $\sigma_{x'} = \sigma_{y'}$ and $\varphi = 0$, corresponding to a circular cloud. Using the fit parameters, we reconstructed both distributions and subtracted the circular fit from the elliptic fit. This subtraction isolates the residual deformation and reveals a characteristic four-lobe pattern with maxima along the elongated axis of the cloud (see Fig.~\ref{fig:fig_2}b).

Second, we rotated the average image by $\ang{90}$ and subtracted it from itself. For an elliptic cloud, this operation again yields a four-lobe pattern. We radially integrated this pattern and fitted it with $\mathrm{OD}(\vartheta) = A\sin(2\vartheta + \varphi)$, where $\vartheta$ was the polar angle in the imaging plane and $\varphi$ gives the orientation of the major axis. Note that the rotation can only be performed in integer pixel steps; the residual pattern is slightly affected by this pixelation, which reduces its apparent symmetry.

\begin{figure*}
    \centering
    \includegraphics[width=\linewidth]{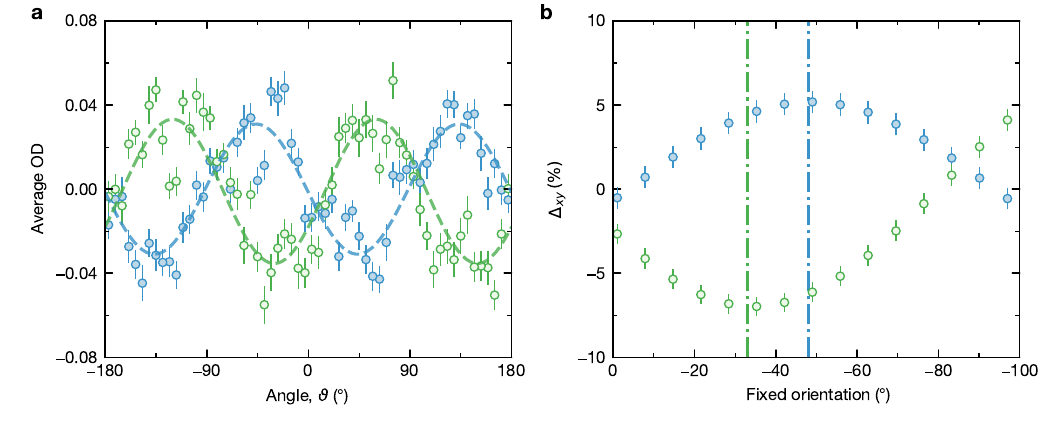}
    \caption{\textbf{Radial integration of self-subtracted image and effect of fixing the coordinate axis for fitting deformation.} \textbf{a)} Blue (Green) points show the radial average of the self-subtracted image for $\xi = \ang{-10}$($+\ang{10}$). We binned our data in 80 different angular coordinates, and each point on the plot shows the averaged optical density in each bin, integrating over the radial coordinate. Error bars are the standard error of the mean of all points in each angular bin. Dashed lines are a sinusoidal fit to the data with a frequency of two; from this, we extracted the major axis orientation of the ellipsoidal cloud. \textbf{b)} Fitted deformation for different fixed orientations of the coordinate axes. The dash-dotted line indicates the angle extracted from fits to the radially integrated images that was used for the deformation presented in Fig.~\ref{fig:fig_2}.}
    \label{fig:extended_6:theta_fit}
\end{figure*}

Both methods yield angles between the major and camera axes that agree within the uncertainty. We show the sensitivity of fitted deformation on the fixed orientation $\varphi$ in Fig.~\ref{fig:extended_6:theta_fit}. Although the gas was in global thermal equilibrium with a single temperature, modeling the distribution of the gas under FSD effect as different $\sigma$'s, rather than an anisotropic chemical potential, captures the deformation while preserving an analytic fit.

\subsection{Theory}\label{sec:theory}

\subsubsection{Mean-field calculation of Fermi surface deformation}
The trapped ultracold molecular gas is described by the Hamiltonian,
\begin{equation}\label{I0}
	\hat{H} = \hat{H}_{\rm K} + \hat{H}_{\rm U} + \hat{H}_{\rm V}.
\end{equation}
Here, $\hat{H}_{\rm K} = -\int d^3\bm{r} \, \hat{\psi}^\dagger(\bm{r}) \, \hbar^2\nabla^2/(2M) \, \hat{\psi}(\bm{r})$ denotes the kinetic energy, with $\hat{\psi}(\bm{r})$ the fermionic field operator and $M$ the mass of the molecule. The single-particle potential energy is given by $\hat{H}_{\rm U} = \int d^3\bm{r} \, U(\bm{r})\hat{\psi}^\dagger(\bm{r})  \hat{\psi}(\bm{r})$, where $U(\bm{r}) = M(\omega_x^2 x^2 + \omega_y^2 y^2 + \omega_z^2 z^2)/2$ describes an anisotropic harmonic trap with trapping frequencies $\omega_{i=x,y,z}$. For simplicity, we assume $\omega_x = \omega_y$ hereafter.
\begin{equation}\label{I0b}
	\hat{H}_{\rm V}=\frac{1}{2}\int d^3rd^3r^{\prime}\hat{\psi}^{\dagger}(\bm r)\hat{\psi}^{\dagger}(\bm r^{\prime})V(\bm{r}-\bm{r^{\prime}})\hat{\psi}(\bm r^{\prime})\hat{\psi}(\bm r)
\end{equation}
denotes the interacting energy. In the presence of double MW fields ($\pi$ and $\sigma^{-}$) and finite ellipticity in $\sigma^{-}$ MW, the effective interaction between molecules takes the form ${V}(\bm r)={V}_0(\bm r)+\Delta{V}(\bm r)$, where ${V}_0(\bm r)= \sqrt{\frac{16\pi}{5}} C_3 \mathrm{Y}_{20}(\theta,\phi)/r^3$ and $\Delta{V}(\bm r)=\sqrt{\frac{8\pi}{15}}\bar{C}_3\left[\mathrm{Y}_{22}(\theta,\phi)+\mathrm{Y}_{2-2}(\theta,\phi)\right]/r^3$ with $C_3$($\bar{C}_3$) 
 \begin{equation}
     \begin{aligned}
C_3&=\sqrt{\frac{15}{2\pi}}\frac{\eta}{48\pi}(3\cos2\beta-1)\cos^2\alpha\sin^2\alpha\\  \bar{C}_3&=\sqrt{\frac{15}{2\pi}}\frac{\eta}{8\pi}\sin^2\alpha \cos^2\alpha \sin2\xi\cos^2\beta    
     \end{aligned}
 \end{equation}
 being the strength of the (non-)cylindrically symmetric part of the interaction, where $\eta =\sqrt{8\pi /15}\,d_0^{2}/\epsilon _{0}$ with $\epsilon _{0}$
being the electric permittivity of vacuum and $d_0$ the magnitude of the permanent electric dipole moment of each molecule. Three Euler angles $\alpha $, $\beta $, and $\gamma $ are determined by the parameters of the MW fields~\cite{Wzhang2025}. 
$\mathrm{Y}_{lm}(\theta,\phi)$ are spherical harmonics with $\theta$ and $\phi$ being the polar and azimuthal angles, respectively. 
 
Following the approach in Ref.~\cite{Zhang2010,Zhang2011}, we apply the mean-field and local density approximations to obtain the thermal equilibrium phase-space distribution function (PSDF) at temperature $T$,
\begin{equation}
    f(\bm{r}, \bm{k}) = \left\{ \exp\left[\frac{\epsilon(\bm{r}, \bm{k}) - \mu}{k_\mathrm{B} T}\right] + 1 \right\}^{-1} \geq 0,
\end{equation}
where $\epsilon(\bm{r}, \bm{k}) = \hbar^2 \bm{k}^2 / (2M) + U_{\mathrm{eff}}(\bm{r}, \bm{k})$ is the quasi-particle energy. The effective potential consists of the trap, Hartree, and Fock contributions: $U_{\mathrm{eff}}(\bm{r}, \bm{k}) = U(\bm{r}) + \epsilon_H(\bm{r}) + \epsilon_F(\bm{r}, \bm{k})$,
with the Hartree term given by $\epsilon_H(\bm{r}) = \int d^3\bm{r}^{\prime} V(\bm{r} - \bm{r}^{\prime}) n(\bm{r}^{\prime})$, and the Fock term by $\epsilon_F(\bm r,\bm k)=-\int d^3k^{\prime}\tilde{V}(\bm{k}-\bm{k^{\prime}})f(\bm r,\bm k^{\prime})/{(2\pi)^3}$.
Here, $n(\bm{r}) = \int d^3\bm{k} \, f(\bm{r}, \bm{k}) / (2\pi)^3$ is the real-space density distribution, and $\tilde{V}(\bm{k})$ is the Fourier transform of the effective interaction. The chemical potential $\mu$ is determined self-consistently by enforcing the total particle number constraint $N\equiv \int d^3rd^3k f(\bm r,\bm k)/(2\pi)^3$.

To avoid significant computational overhead, we combine a self-consistent iteration scheme with a perturbative correction to compute the deformation of the FS. The procedure is implemented as follows. First, we neglect the anisotropic component $\Delta V$ and apply the self-consistent method developed in Ref.~\cite{Zhang2011} to obtain the cylindrically symmetric parts of the distribution of quasiparticle energy, denoted as
 \begin{equation}
 	\begin{aligned}
f_{00}\left(\bm{r}, \bm{k}\right)=f_{00}\left(\rho,z;k_{\rho},k_z\right)
 	\end{aligned}
 \end{equation}
 and
  \begin{equation}
 	\begin{aligned}
\epsilon_{00}(\bm{r}, \bm{k})=\epsilon_{00}(\rho,z;k_{\rho},k_z),
 	\end{aligned}
 \end{equation}
respectively. Here, we introduce column coordinates in real and momentum space: $\bm r\equiv(\rho,z,\phi)$ and $\bm k\equiv(k_\rho,k_z,\phi_k)$, where $\rho$ ($k_\rho$), $z$ ($k_z$) and $\phi$ ($\phi_k$) denote the radial distance, height, and azimuthal angle in real (momentum) space, respectively. Next, we reintroduce $\Delta V$ and evaluate its leading-order corrections to $f$ and $\epsilon$, i.e., $\Delta f(\bm r,\bm k)$ and $\Delta \epsilon(\bm r,\bm k)$. 
A leading-order expansion of the Fermi-Dirac distribution yields,
 \begin{equation}
 	\begin{aligned}
f(\bm r,\bm k)&=\frac{1}{e^{\beta(\epsilon(\bm r,\bm k)-\mu)}+1}=\frac{1}{e^{\beta(\epsilon_{00}+\Delta\epsilon-\mu)}+1}\\
&=f_{00}(\bm r,\bm k)+\Delta f(\bm r,\bm k)\\
&\approx f_{00}[1-\frac{\beta e^{\beta(\epsilon_{00}-\mu)}}{e^{\beta(\epsilon_{00}-\mu)}+1}\Delta\epsilon(\bm r,\bm k)+\mathcal{O}(\Delta\epsilon^2)],\
 	\end{aligned}
 \end{equation}
hence the correction
\begin{equation}
\Delta f(\bm r,\bm k)\approx -\beta e^{\beta[\epsilon_{00}(\bm r,\bm k)-\mu]}f_{00}^2(\bm r,\bm k)\Delta\epsilon(\bm r,\bm k), \end{equation}
where $\beta=1/(k_\mathrm{B}T)$.

Similarly, the quasi-particle energy $\epsilon(\bm r,\bm k)=\epsilon_{00}(\bm r,\bm k)+\Delta\epsilon(\bm r,\bm k)$, where
\begin{equation}
 	\begin{aligned}
\Delta\epsilon(\bm r,\bm k)&\approx\int d^3r^{\prime}\Delta V(\bm{r}-\bm{r^{\prime}})n_{00}(\bm r^{\prime})\\
&-\int \frac{d^3k^{\prime}}{(2\pi)^3}\Delta\tilde{V}(\bm{k}-\bm{k^{\prime}})f_{00}(\bm r,\bm k^{\prime})\\
&+\int d^3r^{\prime}V_0(\bm{r}-\bm{r^{\prime}})\Delta n(\bm r^{\prime})\\
&-\int \frac{d^3k^{\prime}}{(2\pi)^3}\tilde{V}_0(\bm{k}-\bm{k^{\prime}})\Delta f(\bm r,\bm k^{\prime}).
 	\end{aligned}
 \end{equation}
Here, 
\begin{equation}
    \begin{aligned}
n_{00}(\bm r)\equiv\int \frac{d^3k}{(2\pi)^3} f_{00}(\bm r,\bm k)        
    \end{aligned}
\end{equation}
and
\begin{equation}
    \begin{aligned}
\Delta n(\bm r)\equiv\int \frac{d^3k}{(2\pi)^3} \Delta f(\bm r,\bm k)        
    \end{aligned}
\end{equation}
are the density distribution and its first-order correction, respectively.

Since $\Delta V(\bm{r}) \propto \cos(2\phi)$ and $\Delta\tilde{V}(\bm{k}) \propto \cos(2\phi_k)$, the ansatz
\begin{equation}
\begin{aligned}
\Delta f(\bm{r}, \bm{k}) = &\Delta f_{20}(\rho,z;k_\rho,k_z)\cos(2\phi) \\
&+ \Delta f_{02}(\rho,z;k_\rho,k_z)\cos(2\phi_k)
\end{aligned}
\end{equation}
and
\begin{equation}
\begin{aligned}
\Delta\epsilon(\bm{r}, \bm{k}) = &\Delta\epsilon_{20}(\rho,z;k_\rho,k_z)\cos(2\phi)\\
&+ \Delta\epsilon_{02}(\rho,z;k_\rho,k_z)\cos(2\phi_k)
\end{aligned}
\end{equation}
provide a self-consistent solution.
Finally, the FSD is extracted based on $f_{00}$ and its perturbative correction. 
The momentum distribution is given by,
 \begin{equation}
 	\begin{aligned}
\tilde{n}(\bm k)&\equiv\int d^3r[f_{00}(\bm r,\bm k)+\Delta f(\bm r,\bm k)],
 	\end{aligned}
\end{equation}
which determines the momentum-space widths $\sigma_{x,y}\equiv[\int{d^3k}\tilde{n}(\bm k)k_{x,y}^2/{(2\pi)^3}]^{1/2}$. In the TOF experiment, the aspect ratio $\Delta_{xy} \equiv{\sigma_x}/{\sigma_y}-1$ is measured.

\subsubsection{Calculation of Dipolar-to-Fermi energy ratio}\label{sec:comparisons}

To quantify the system's proximity to the strong interaction regime, we evaluate the ratio of the dipolar interaction energy to the Fermi energy, $E_{\mathrm{dd}}/E_{\mathrm{F}}$. The Dipolar energy is obtained from the dipolar length $a_{\text{dd}}$ as,
\begin{equation}
E_{\text{dd}} = \frac{n_0 d_\text{eff}^2}{4\pi\epsilon_0} = \frac{n_0 2 \hbar^2 a_{\text{dd}}}{M},
\label{equation:E_dd_formula}
\end{equation}
where $d_\text{eff}$ is the the effective dipole moment, $M$ is the mass of the dipoles,   and $n_0$ is the peak density of the gas. Within the local density approximation (LDA), the trapped gas is treated as locally homogeneous at the trap center. Accordingly, we approximate the density entering Eq.~\ref{equation:E_dd_formula} by the peak density.

For a harmonically trapped Fermi gas at finite temperature, the peak density is expressed in terms of the fugacity $\zeta$ as, 
\begin{equation}
n_{0}(M,T,\zeta) = -\big(\frac{M k_B T}{2\pi \hbar^2}\big)^{3/2}
\,\mathrm{Li}_{\frac{3}{2}}(-\zeta),
\end{equation}
where $\mathrm{Li}_{\nu}$ denotes the polylogarithm of order $\nu$. The fugacity $\zeta$ is extracted by fitting the measured density distributions with the finite-temperature Fermi-Dirac profile (see Eq.~\ref{equation:fermi_dirac}), and the temperature $T$ is obtained via $T = T_\mathrm{F}\left[-\frac{1}{6\,\mathrm{Li}_3(-\zeta)}\right]^{1/3}$. Here, the Fermi temperature for a harmonic trapped gas is given by $T_\mathrm{F} = E_\mathrm{F}/k_\mathrm{B} = \hbar\bar{\omega} (6N)^{1/3}/k_\mathrm{B}$, where $\bar{\omega}$ is the geometric mean trapping frequency and $N$ is the total particle number. 

Using these relations, we obtain $n_0 \approx 2.28\times 10^{12}\,\text{cm}^{-3}$, corresponding to $E_{\text{dd}} = k_\mathrm{B}\times 5.5 \, \text{nK}$ at $3000\, a_0$ dipolar length. With a Fermi energy of $E_{\text{F}} = k_\mathrm{B}\times120\, \text{nK}$, this yields $E_{\mathrm{dd}}/E_{\mathrm{F}} = 0.046$, approximately a factor of five larger than that reported for magnetic erbium atoms~\cite{Aikawa2014}.

\end{document}